\documentclass[conference,10pt]{IEEEtran}

\IEEEoverridecommandlockouts

\usepackage{graphicx}
\usepackage{algorithm}
\usepackage{algpseudocode}
\usepackage{pifont}
\usepackage{amsmath}
\usepackage{float}

\usepackage{mwe}
\usepackage{subfig}
\usepackage{epstopdf}
\usepackage[T1]{fontenc}
\usepackage[utf8]{inputenc}
\usepackage[affil-it]{authblk}
\DeclareGraphicsExtensions{.eps}
\usepackage[font=scriptsize]{caption}

\usepackage{array}
\usepackage{amsmath}
\usepackage{mathtools}

\DeclareMathOperator*{\argmax}{argmax}

\makeatletter
\g@addto@macro\normalsize{%
  \setlength\abovedisplayskip{3pt}
  \setlength\belowdisplayskip{3pt}
 \setlength\abovedisplayshortskip{3pt}
\setlength\belowdisplayshortskip{3pt}
}
\makeatother

\begin{document}
\bstctlcite{IEEEexample:BSTcontrol}
\title{On the Number and 3D Placement of Drone Base Stations in Wireless Cellular Networks}

\author[*]{Elham Kalantari}
\author[**]{Halim Yanikomeroglu}
\author[*]{Abbas Yongacoglu}
\affil[*]{School of Electrical Engineering and Computer Science, University of Ottawa, Ottawa, ON, Canada}
\affil[**]{Department of Systems and Computer Engineering, Carleton University, Ottawa, ON, Canada}

\renewcommand\Authands{, and }

\maketitle

\begin{abstract}
Using drone base stations (drone-BSs) in wireless networks has started attracting attention. Drone-BSs can assist the ground BSs in both capacity and coverage enhancement. One of the important problems about integrating drone-BSs to cellular networks is the management of their placement to satisfy the dynamic system requirements. In this paper, we propose a method to find the positions of drone-BSs in an area with different user densities using a heuristic algorithm. The goal is to find the minimum number of drone-BSs and their 3D placement so that all the users are served. Our simulation results show that the proposed approach can satisfy the quality-of-service requirements of the network.
\end{abstract}


\IEEEpeerreviewmaketitle

\section{Introduction}
\begin{figure}[t]
  \begin{center}
  \includegraphics[width=3.2in]{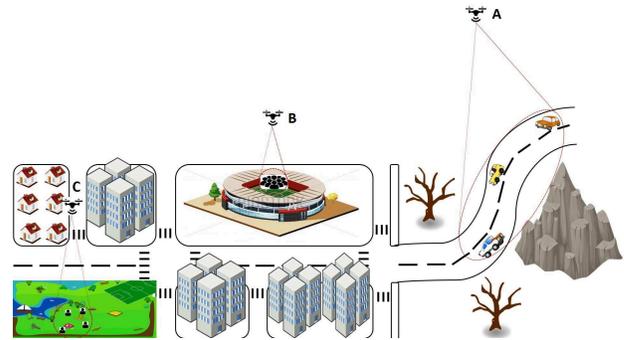}\\
  \end{center}
  \caption{Drone-BSs can tackle coverage (A or C) or capacity (B or C) issues.}
  \label{drone}
\end{figure}

Wireless users expect unlimited capacity everywhere and all the time, at an affordable price. The brute-force way to provide ubiquitous high-rate coverage is very well known: deploy a very dense network of BSs. However, this solution is not feasible in terms of CapEx and OpEx, due to the fact that a high percentage of these BSs will be lightly loaded or even will not have any load at a given time and space. Moreover, the temporal and spatial variations in user densities and user application rates are expected to result in difficult-to-predict traffic patterns. The utilization of drone-BSs is a promising solution in such scenarios. \\
\indent Drone-BSs can help the ground network of BSs in providing high data rate coverage whenever there is an excessive need in space and time, especially in situations when this excessive demand occurs in a rather difficult-to-predict manner (Fig. \ref{drone}). Although the OpEx of a drone-BS may be more than that of a ground small-cell BS, if engineered properly, drone-BSs may result in substantial savings in the overall network deployment costs.\\
\indent There are a growing number of papers related to drone base stations in cellular networks. In \cite{7037248} air-to-ground pathloss for low altitude platforms (LAP), like drone-BSs with heights less than 3000 meters is modelled. The model shows that there are two main propagation groups, corresponding to the receivers with line-of-sight (LoS) connections and the ones without LoS connections which still receive the signal from LAP due to strong reflections and diffractions. In \cite{6863654} a closed form expression for the probability of LoS connection between a LAP and a receiver is developed, and then through an analytical approach the optimum altitude that maximizes the radio coverage is obtained. In \cite{MozaffariSBD15} adding an unmanned aerial vehicle (UAV) to a device-to-device (D2D) network is investigated and the average coverage probability and the sum-rate for the users is derived as a function of UAV altitude and number of D2D users. In \cite{MozaffariSBD15a} the optimum altitude of a single drone-BS to obtain a required coverage while minimizing the transmit power is found. Also providing maximum coverage with two drone-BSs in the presence and absence of interference is investigated. In \cite{Irem} the authors propose using a drone-BS to maximize the revenue of the network. They formulate a 3D problem with the objective of maximizing the number of users that are under the coverage of a drone-BS and find the 3D placement of the drone-BS through numerical methods.\\
\indent A very important question about the aerial wireless networks, which has not been addressed yet in the literature, is to find the minimum number of drone-BSs along with their placement in order to provide coverage to a set of users with some target Quality-of-Service (QoS).  The main contribution of this paper is to provide a heuristic algorithm to find the minimum number of drone-BSs and their 3D locations to serve an arbitrarily located set of users. (In this paper, we assume that there are no ground BSs; in an extended work, we plan to consider both aerial and grounds BSs.)\\
\indent The rest of this paper is organized as follows. In Section II the system model is presented. The proposed placement algorithm is described in Section III. Simulation results are given in Section IV, followed by conclusions in Section V.

\section{System Model}

Consider an area with a specific number of users. Our goal is to find the minimum number of drone base stations and their 3D placement to give service to the users in the region. We limit our analysis to downlink, so drone-BSs are transmitting data. An important feature of a drone-BS is its ability to move, therefore we do not need to cover a region while there is no user there. As users move, drone-BSs might follow them if needed, so here we find the placement of the drone-BSs for one snapshot of the users positions.

\subsection{Air-to-ground channel model}
\begin{figure}[t]
	\begin{center}
		\includegraphics[width=3.4in]{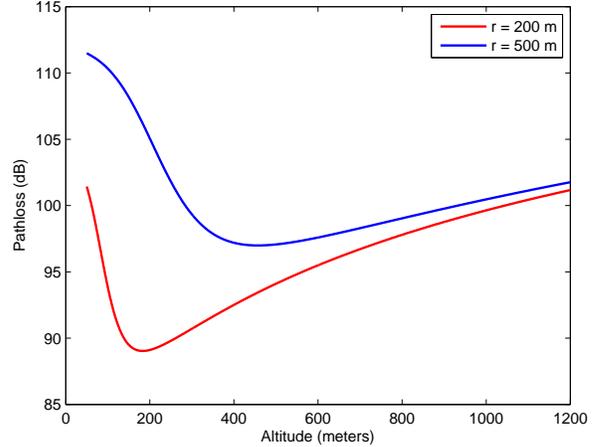}\\ 
	\end{center}
	\caption{Pathloss versus the altitude of a drone-BS for two fixed horizontal distances.}
	\label{fig1}
\end{figure}

The air-to-ground channel model is different from terrestrial channel models. The probability of LoS is an important factor in modelling air-to-ground pathloss. This probability can be formulated as \cite{6863654}
\begin{equation}\label{LoS}
P(\textrm{LoS}) = \frac{1}{1+a \exp (-b(\frac{180}{\pi}\theta -a))},
\end{equation}
where $a$ and $b$ are constant values depending on the environment (rural, urban, etc) and $\theta$ is the elevation angle equal to $\arctan (\frac{h}{r})$, where $h$ and $r$ are the altitude of a drone-BS and its horizontal distance from the receiver, respectively. Equation (\ref{LoS}) shows that the probability of having LoS connection is increased as the elevation angle increases, so if $r$ is fixed by increasing the altitude of a drone-BS, the probability of LoS will increase. Here we do not consider the random behaviours of the radio channel and the mean pathloss model we adopt is the one given in \cite{6863654}
\begin{eqnarray}
\nonumber
\textsf{PL}(\textrm{dB}) &=& 20 \log(\frac{4\pi f_c d}{c}) \\
&+&P(\textrm{LoS})\eta_{LoS}+P(\textrm{NLoS})\eta_{NLoS},
\end{eqnarray}
where $f_c$ is the carrier frequency, $c$ is the speed of light, $d$ is the distance between a drone-BS and the receiver and is equal to $\sqrt{h^2+r^2}$. $P(\textrm{NLoS})=1-P(\textrm{LoS})$, $\eta_{LoS}$ and $\eta_{NLoS}$ are average additional loss to the free space propagation for LoS and NLoS connection, respectively, depending on the environment.
The air-to-ground pathloss versus altitude in an urban environment for $r=200$ and $r=500$ meters and $f_c=2$ GHz is shown in Fig.~\ref{fig1}. As it is seen in this figure, by increasing the altitude of a drone-BS, the pathloss first decreases and then increases. That is because in low altitudes the probability of NLoS is much higher than that of LoS, due to reflections by buildings and other objects, and the additional loss of a NLoS connection is higher than a LoS connection; but when the altitude increases the LoS probability increases as well and in turn path loss decreases. On the other hand, the pathloss is also dependent on the distance between the transmitter and the receiver, so after a specific height this factor dominates and as the altitude increases, the pathloss increases as well. The drone base stations can be considered as a new tier of access nodes in cellular communication systems. Instead of changing the transmit power, which makes different coverage areas, the desired coverage area can be attainable by changing the height of a drone-BS. 

\subsection{Optimization problem}
\indent We adopted the framework of our optimization problem from \cite{7056465}. To start the optimization problem, we need an initial estimation of the number of drone-BSs that can serve all the users. The number of drone-BSs should be estimated based on both coverage and capacity requirements. For coverage constraint, according to \cite{6863654} only one drone-BS can cover a very large area by setting its altitude in the optimum height, so in our problem mainly the data traffic requirement of the users enforces the usage of more drone-BSs. Therefore for the initial estimation of the number of drone-BSs we only consider the capacity constraint. First, we need to find the maximum number of users that a drone-BS can serve
\begin{equation}
N_{U_{BS}} = \lfloor  \frac{C_{BS}}{R}  \rfloor,
\end{equation}
where $\lfloor .\rfloor$ denotes the floor function, $R$ is the target download rate of the users, and $C_{BS}$ is the capacity of a drone-BS and is equal to $B\times \eta$, where $B$ is the total bandwidth of the drone-BS and $\eta$ is the average spectral efficiency of the system. The number of drone-BSs is estimated as
\begin{equation}
N_{BS} = \lceil  \frac{N_{U}}{N_{U_{BS}}}  \rceil,
\end{equation}
where $\lceil .\rceil$ denotes the ceiling function and $N_{U}$ is the total number of users available in the region.\\
\indent Our goal, as it is mentioned earlier, is to cover the users and serve them based on their traffic requirements. For the coverage constraint, we wish at least $\zeta$ percent of all the users to be covered by drone-BSs. This constraint can be formulated as
\begin{equation}
P \{ \sigma_{ij^*}>\gamma_{th} \} \ge \zeta,   i=1,...,N_{U},
\end{equation}
where $j^* = \argmax\limits_{j} \sigma_{ij}$, and $\sigma_{ij}$ is the SINR of user $i$ receiving service from drone-BS $j$, and $\gamma_{th}$ is the minimum SINR level required for each user.\\
\indent For the capacity constraint, a parameter $\rho_{j,k}$ ($j=1,...,N_{BS} , \text{ and } k=1,...,N_{subarea}$ ) is defined as \cite{7056465}
\begin{equation}
\rho_{j,k} = \frac{a_{j,k}}{A_j},
\end{equation}
where $a_{j,k}$ is the mutual area between drone-BS $j$ and subarea $k$, and $A_j$ is the total area of drone-BS $j$; so $\rho_{j,k}$ is between zero and one. A subarea is part of the region in which the user density is the same.
To satisfy the capacity constraint the below inequality should hold:
\begin{equation}\label{cap}
\sum_{j=1}^{N_{BS}}N_{U_{BS}}\rho_{j,k} \ge D_k S_k , k=1,...,N_{subarea},
\end{equation}
where $D_k$ and $S_k$ are the user density function and the area of subarea $k$, respectively. \\
\indent Finally the optimization problem is expressed as follows:
\begin{equation}
 \underset{x_j,y_j,h_j,\{\epsilon_j\}}{\text{minimize}} \sum_{j=1}^{N_{BS}} \epsilon_j 
\end{equation}
\indent subject to:
\begin{eqnarray}
 \sum_{j=1}^{N_{BS}}N_{U_{BS}}\rho_{j,k} &\ge& D_k S_k , k=1,\ldots,N_{subarea} \\
 \sum_{i=1}^{N_U}I_i &\ge& \zeta N_U\\
 \frac{1}{\textsf{E}\{\frac{1}{\eta_i}\}} &\ge& \eta,\label{SE}
\end{eqnarray}

where (\ref{SE}) ensures that the average spectral efficiency of the system is at least $\eta$ (the value we used in initial estimation part), and $\eta_i$ is the spectral efficiency of user $i$. Also $x_j$, $y_j$, and $h_j$ are 3D positions of the drone-BS $j$. We also define the following indicator functions:
\begin{equation}
\epsilon_j=
\begin{dcases}
 1, \text{ if drone-BS } j \text{ is used,} \\
0, \text{ if drone-BS } j \text{ is redundant,}
\end{dcases}
\end{equation}
\begin{equation}
I_i=
\begin{dcases}
 1, \text{ if user } i \text{ is under the coverage of a drone-BS,} \\
0, \text{ otherwise.}
\end{dcases}
\end{equation}

\section{proposed algorithm}

\begin{algorithm}[tb]
\caption{PSO algorithm for 3D placement of drone-BSs}
\label{PSOalgorithm}
\begin{algorithmic}[1]
\State Generate an initial population including $L$ random particles $W^{(l)}(0),l=1,...,L$. Each particle has size $3\times N_{BS}$.
Set $t=1$, $U=U_1$, $U^{(global)}=\min \{{U^{(l)}(0), l=1,...,L} \}$ and $U^{(l,local)}=U^{(l)}(0)$.
\While {$U^{(global)}>-N_U$ }
\For {l = 1, ..., L}
\State Compute $V^{(l)}(t), W^{(l)}(t), U^{(l)}(t)$.
\If {$U^{(l)}(t)<U^{(l,local)}$} 
\State $W^{(l,local)}=W^{(l)}(t), U^{(l,local)}=U^{(l)}(t)$. 
\If {$U^{(l,local)}<U^{(global)}$}
\State $W^{(global)}=W^{(l,local)}, U^{(global)}=U^{(l,local)}$. 
\EndIf
\EndIf
\EndFor
\If {$U^{(global)} \le0$}
\State $U=U_2$.
\EndIf
\If {$U^{(global)} \le-\zeta N_U$}
\State $U=U3$.
\EndIf
\State $t=t+1$.
\EndWhile
\end{algorithmic}
\end{algorithm}

Finding BS locations ensuring that that nearly all the users are served and their traffic requirements is achieved is a very complicated problem in general. Adding a new dimension to the problem, which is the altitude of the aerial base stations, makes the problem even more complex. Meta heuristic algorithms such as genetic algorithm \cite{goldberg1989genetic}, simulated annealing \cite{671}, particle swarm optimization \cite{488968}, tabu search \cite{Glover:1997:TS:549765}, and ant colony optimization \cite{484436} are often used in such complex problems. Here we use particle swarm optimization (PSO) algorithm to find the 3D placement of the drone-BSs. The algorithm is an appropriately modified version of the one developed in \cite{7056465} to fit our problem.\\
\indent PSO is an optimization technique proposed by J. Kennedy and R. Eberhart in 1995 \cite{488968}. It is inspired by social behaviour of bird flocking or fish schooling. The algorithm starts with a population of random solutions and iteratively tries to improve the candidate solutions with regards to a given measure of quality. The best experience of each candidate as well as the best global experience of all the candidates in all iterations are recorded and the next movement of the candidates is influenced by these items.\\
\indent In order to find the 3D placement of the drone-BSs, using PSO algorithm, we first consider the capacity constraint and find the locations of drone-BSs that minimize the below utility function:
\begin{equation}
U_1 = \sum_{k=1}^{N_{subarea}} \sum_{j=1}^{N_{BS}}\{ N_{U_{BS}}\rho_{j,k}- D_k S_k \}.
\end{equation}
\indent Then we use these 3D points as an initial solution and improve them so that the total number of uncovered users is minimized, taking into account that the capacity constraint should still hold. The utility function that satisfies coverage constraint, while keeping the capacity constraint active, is given below:
\begin{equation}
  U_2 =  
\begin{dcases}
     - \sum_{i=1}^{N_U}I_i,& \text{if (} \ref{cap} \text{) holds,}\\
    0,              & \text{otherwise.}
\end{dcases}
\end{equation}
\indent Finally, to satisfy (\ref{SE}), we use the following utility function:
\begin{equation}
  U_3 =  
\begin{dcases}
     -N_U+\eta-\frac{1}{\textsf{E}\{\frac{1}{\eta_i}\}}                  ,& \text{if (} \ref{cap} \text{) holds,}\\
    0,              & \text{otherwise.}
\end{dcases}
\end{equation}\\
\indent The PSO algorithm starts by generating $L$ particles of length  $3\times N_{BS}$  to form an initial population $W^{l},l=1,...,L$. Each particle contains random positions of all the drone-BSs within the region. The particle that provides the best utility in all the iterations is recorded as $W^{(global)}$. Also for each particle the best result is kept as $W^{(l,local)}$. In each iteration, $W^{(global)}$ and $W^{(l,local)}$ are updated and the velocity and movement of the particles are calculated based on them. The velocity term $V_w^{(l)}, w=1,...,3N_{BS}$, at iteration $t$ is computed as follows:
\begin{eqnarray}
\nonumber
V_w^{(l)}(t+1)&=&\phi V_w^{(l)}(t)\\
\nonumber
&+&c_1\phi_1(W_w^{(l,local)}(t)-W_w^{(l)}(t)) \\
&+&c_2\phi_2(W_w^{(global)}(t)-W_w^{(l)}(t)), 
\end{eqnarray}
where $\phi$ is the inertia weight that controls speed of convergence. $c_1$ and $c_2$ are personal and global learning coefficients, and $\phi_1$ and $\phi_2$ are two random positive numbers. Afterwards, the positions of the elements in a particle are updated as:
\begin{equation}
W_w^{(l)}(t+1)=W_w^{(l)}(t)+V_w^{(l)}(t+1).
\end{equation}
\indent Details of the proposed algorithm is provided in Algorithm~\ref{PSOalgorithm}. This algorithm finds a feasible solution for drone-BS placement in the region so that $\zeta$ percent of the users are served based on their data traffic requirements. After finding the 3D placement of the drone-BSs, we try to minimize their number by removing the ones whose elimination do not affect the quality of the network. It can iteratively be checked by removing one drone-BS in each iteration and then checking the constraints. If they hold, the drone-BS can be removed without violating the constraints. If more than one drone-BS can be removed based on this approach, at first step the one which results in fewer users disconnected from the system is selected as the redundant drone-BS. After removing this drone-BS, the algorithm is repeated until finally no redundant drone-BS remains \cite{7056465}.

\section{simulation results}
\begin{figure}[tb]

  \subfloat[]{\label{BSC_deploy_uniforma}\includegraphics[width=3.4in]{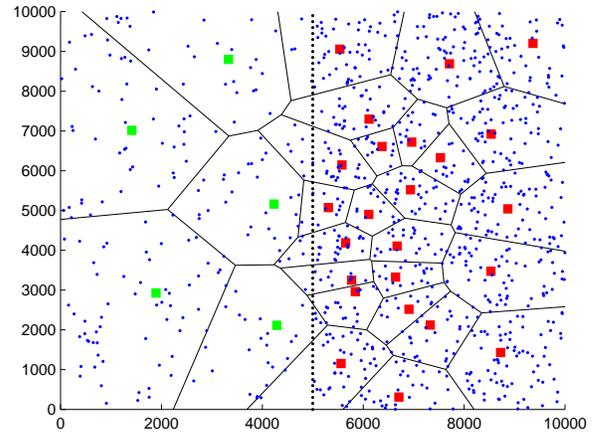}}
  
  \subfloat[]{\label{BSC_deploy_uniformb}\includegraphics[width=3.4in]{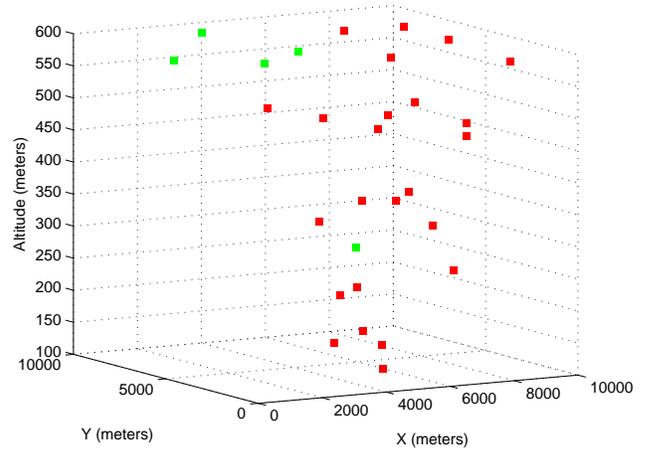}}

   \caption {Scenario I, a) User distribution and 2D projection of drone-BSs locations, b) 3D locations of drone-BSs. Users are uniformly distributed with different densities in the left and right regions. Drone-BSs in the left and right regions are shown by green and red squares, respectively. Users are illustrated by blue dots.}
    \label{BSC_deploy_uniform}
\end{figure}

\begin{table}[t]
\caption{Urban Environment Parameters (left) and Simulation Parameters (right)}\label{table1} 
\parbox[t]{.3\linewidth}{
\centering
\begin{tabular}[t]{|c |c |} 
 \hline
 Parameter & Value \\  
 \hline
 $a$ & 9.61  \\ 
 \hline
 $b$ & 0.16  \\
 \hline
$\eta_{LoS}$ & 1  \\
 \hline
 $\eta_{NLoS}$ & 20 \\
 \hline
\end{tabular}
}
\hfill
\parbox[t]{.7\linewidth}{
\centering
\begin{tabular}[t]{|c |c |c |c |} 
 \hline
 Parameter & Value & Parameter & Value \\ 
 \hline
 $f_c$ & 2 GHz  & $R$ & $1$ Mbps \\ 
 \hline
 $B$ & 20 MHz & $P_t$ & $5$ watts  \\
 \hline
 $\eta$ & $1.7$ bps/Hz &  $\gamma_{th}$ & -7 dB\\
 \hline
 $\zeta$ & 95 & -&-  \\
 \hline
\end{tabular}
}
\end{table}
\begin{figure}[t]
  \begin{center}
  \includegraphics[width=3.4in]{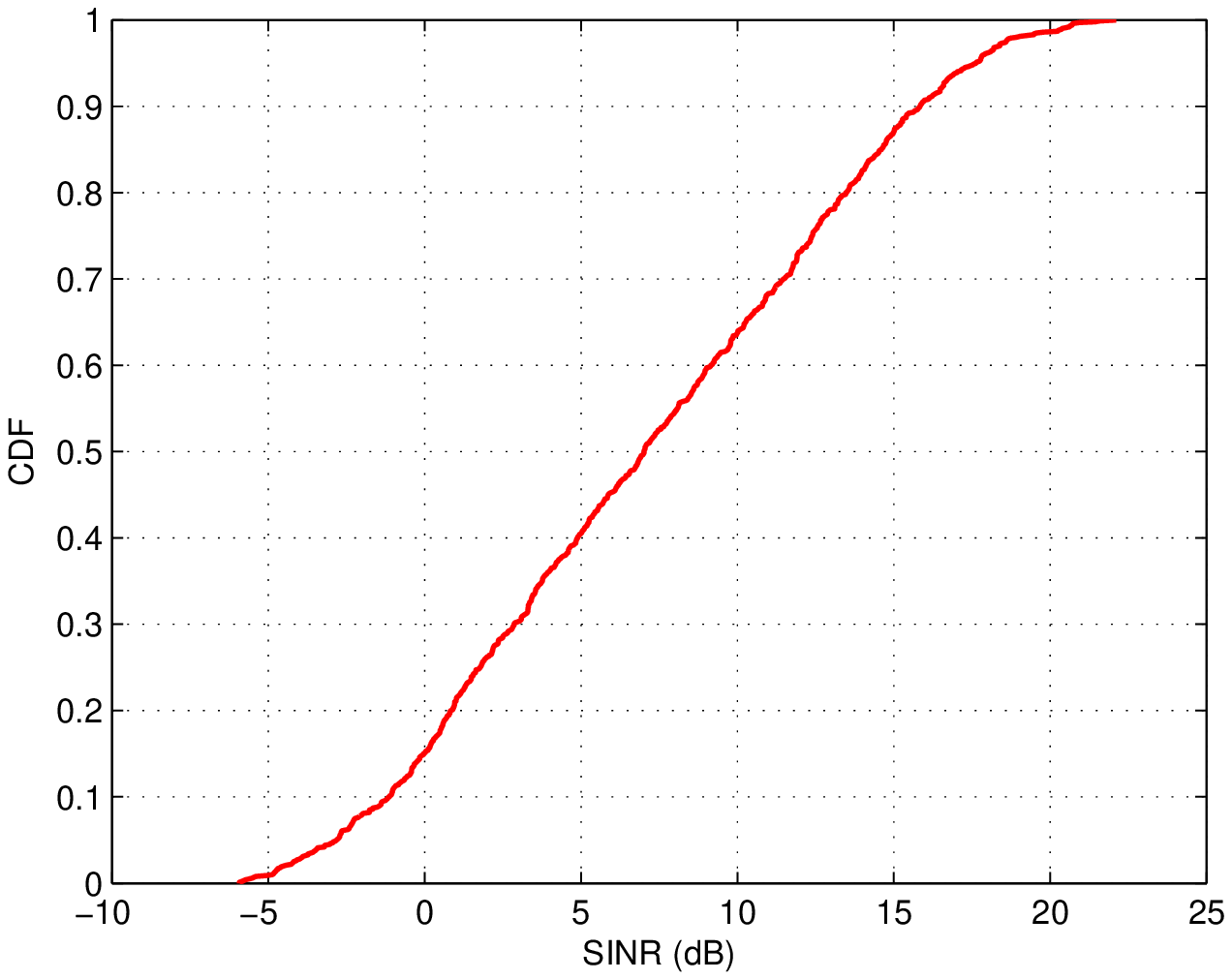}\\
  \end{center}
  \caption{SINR distribution in Scenario I (refer to Fig. \ref{BSC_deploy_uniform}).}
  \label{SINR_uniform}
\end{figure}
\begin{figure}[t]
  \begin{center}
  \includegraphics[width=3.4in]{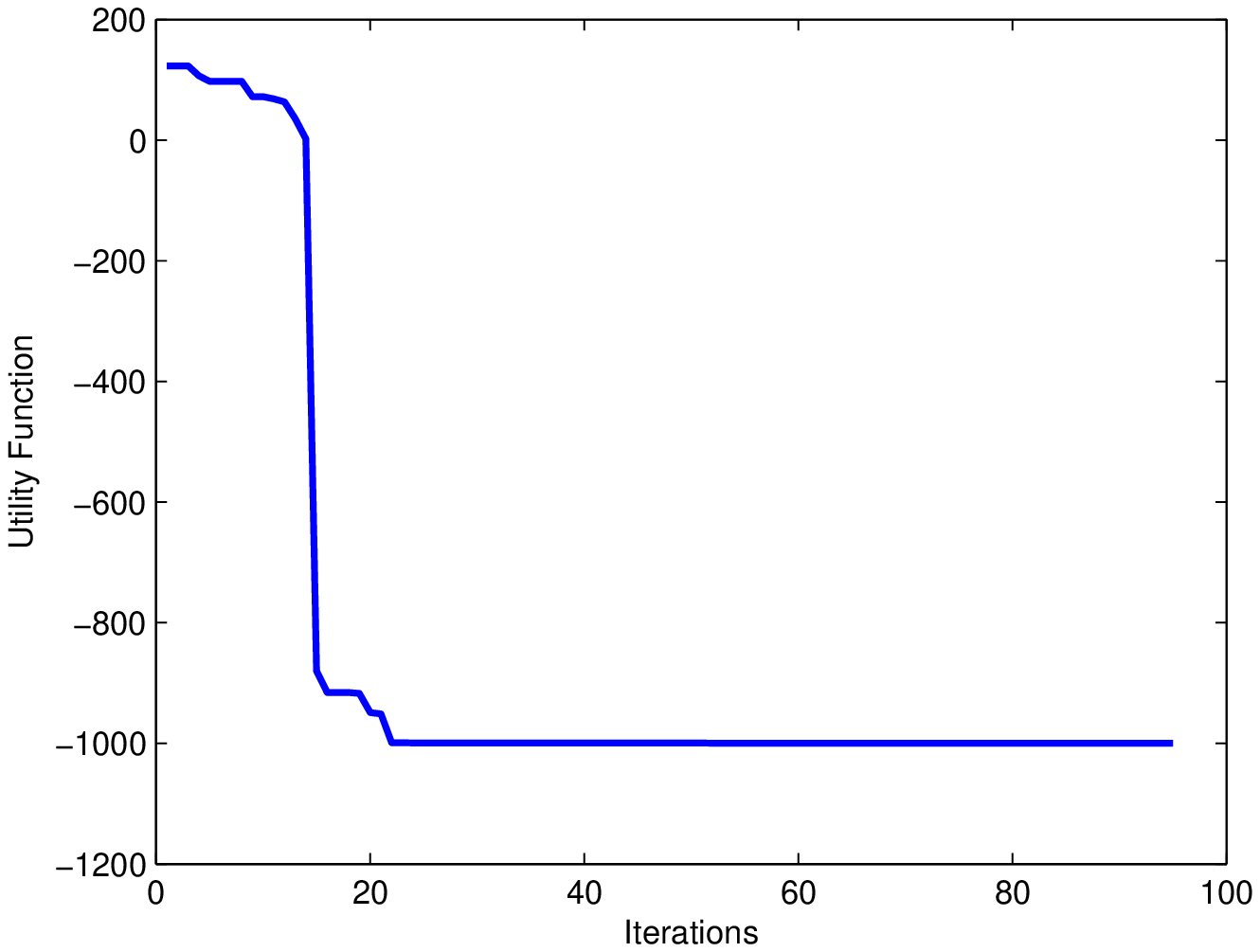}\\
  \end{center}
  \caption{Convergence speed of the PSO algorithm in Scenario I (refer to Fig. \ref{BSC_deploy_uniform}).}
  \label{Iterations_uniform}
\end{figure}

\begin{figure}[t]

  \subfloat[]{\label{BSC_deploy_normala}\includegraphics[width=3.4in]{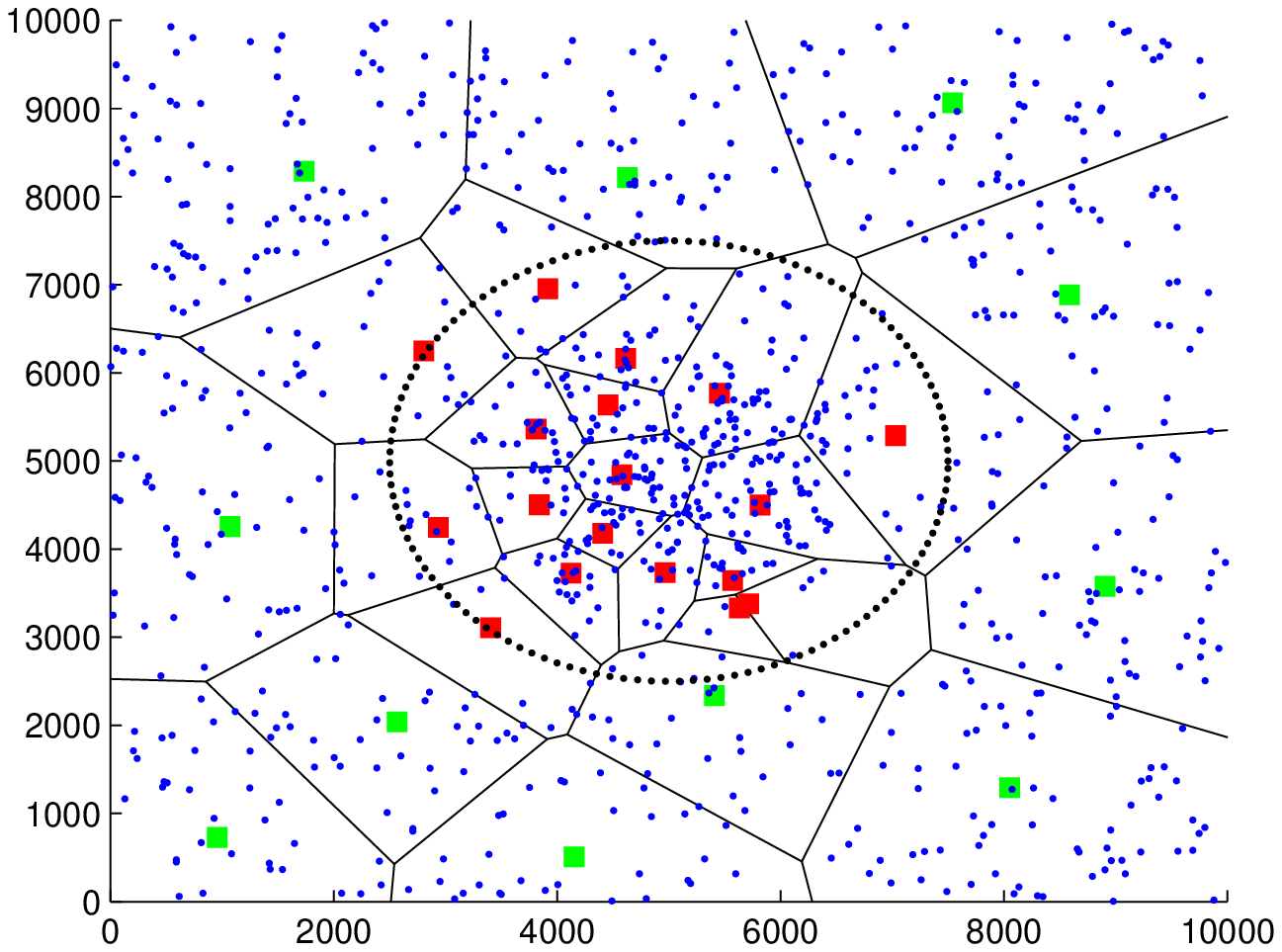}}
  
  \subfloat[]{\label{BSC_deploy_normalb}\includegraphics[width=3.4in]{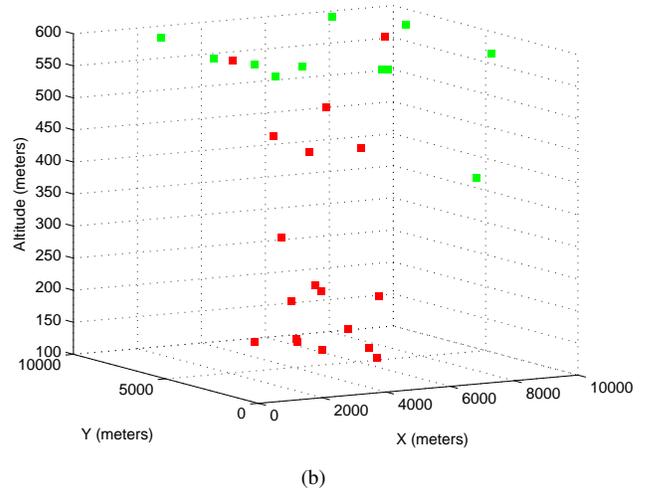}}

   \caption{Scenario II, a) User distribution and 2D projection of drone-BSs locations, b) 3D locations of drone-BSs. 40 percent of users have a normal distribution in the central region and 60 percent of them are uniformly distributed in the remaining region. Drone-BSs in the centeral region and in the remaining region are shown by red and green squares, respectively. Users are illustrated by blue dots.}
   
    \label{BSC_deploy_normal}
\end{figure}
\begin{figure}[t]
  \begin{center}
  \includegraphics[width=3.4in]{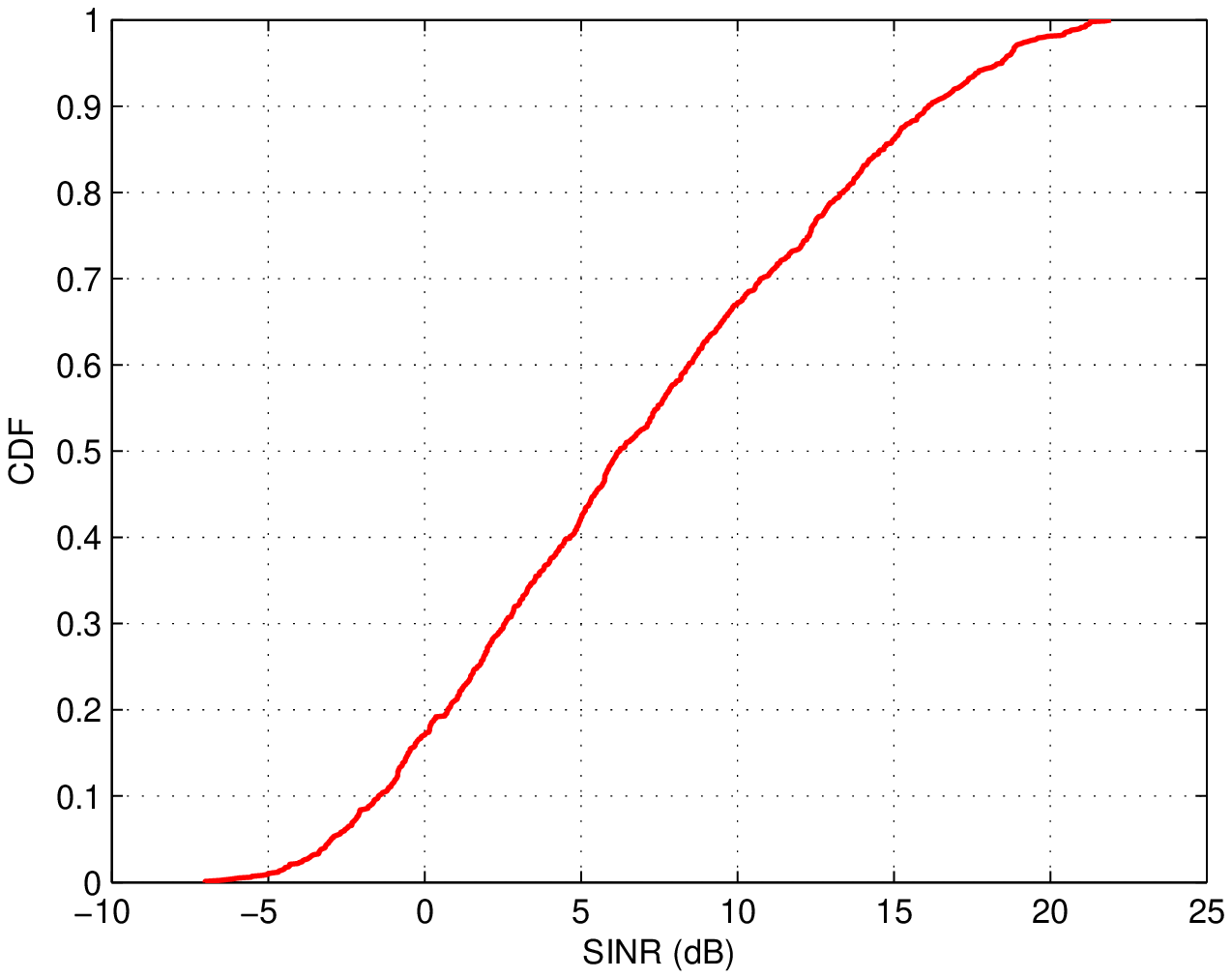}\\
  \end{center}
  \caption{SINR distribution in Scenario II (refer to Fig. \ref{BSC_deploy_normal}). }
  \label{SINR_normal}
\end{figure}
\begin{figure}[t]
  \begin{center}
  \includegraphics[width=3.4in]{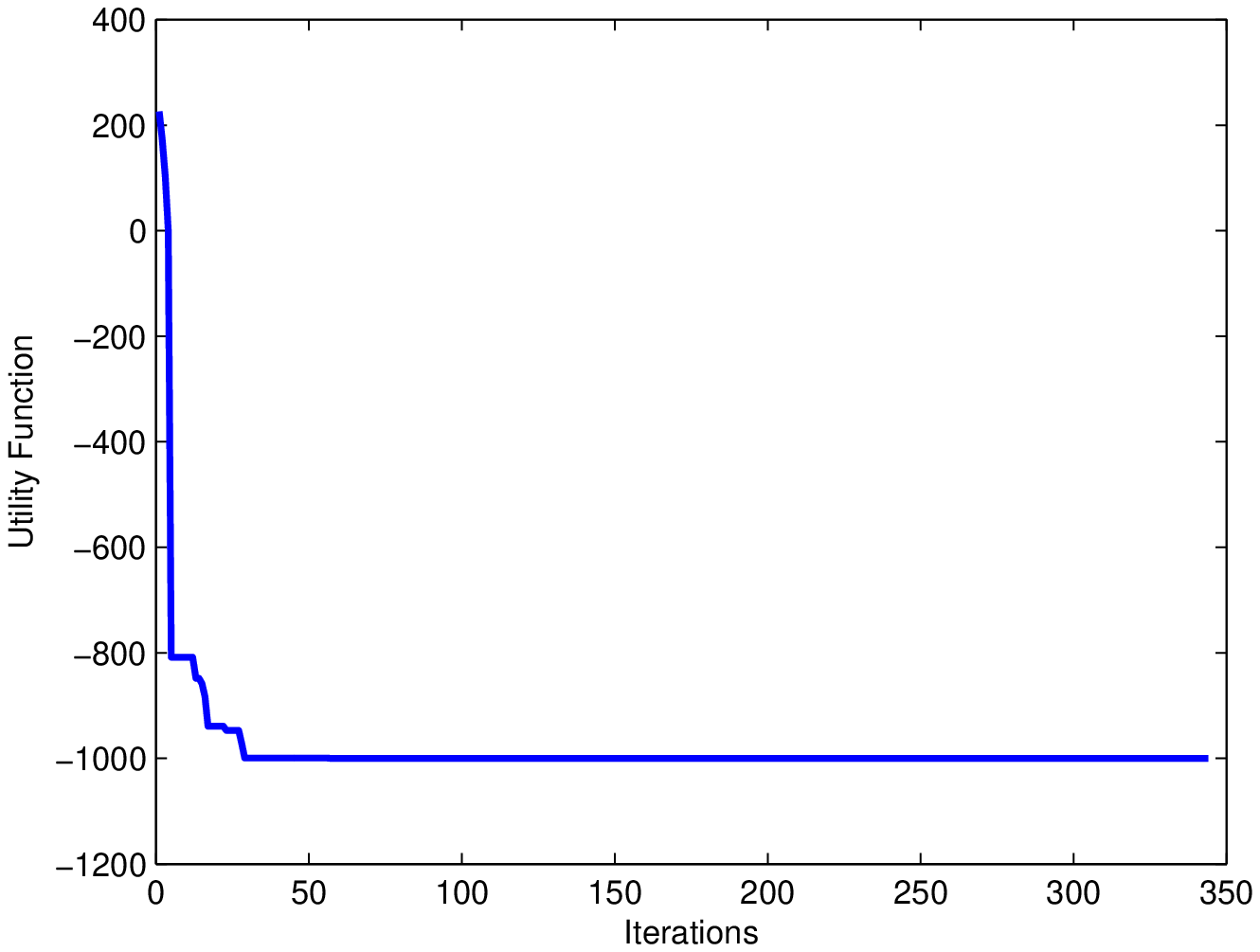}\\
  \end{center}
  \caption{Convergence speed of the PSO algorithm in Scenario II (refer to Fig. \ref{BSC_deploy_normal}). }
  \label{Iterations_normal}
\end{figure}
\indent We consider an urban area with system parameters provided in Table~\ref{table1}. We use Matlab software as a simulation platform. The total area is 100 km$^2$; 1000 users are distributed in this area in a number of different ways. In Scenario I, the total area is divided in two equal regions with 20 and 80 percent of the users uniformly distributed in the left and right regions, respectively. The initial estimation of the number of drone-BSs is 30; finally only 1 drone-BS identified as redundant. The user distribution in this scenario and the 2D projection of drone-BSs locations, using PSO algorithm, are shown in Fig.~\ref{BSC_deploy_uniforma}. As it is seen in this figure, the right region which has higher user density needs more drone-BSs to serve all the users and to avoid congestion. The Voronoi tessellation of the drone-BSs are also shown to give a better insight about the drone-BS placement. It should be noted that these lines do not show the actual frontiers for BS-user connection; the association policy is based on the best SINR. In the PSO algorithm, we limit the maximum altitude of drone-BSs to 600 meters. The 3D placement of the drone-BSs in Scenario I is depicted in Fig.~\ref{BSC_deploy_uniformb}. As seen in this  figure, in the right region which has higher user density, the average altitude of the drone-BSs is less than those in the left region. This is due to the fact that when the region is dense, all the resources of a drone-BS are utilized by the nearby users, as such the drone-BS usually can not serve farther away users. Hence it is better that such a drone-BS decreases its altitude to make less interference to farther users which are served by another drone-BS. In the left region where the user density is lower, the drone-BSs increase their heights to decrease pathloss and cover more users in the region. The CDF curve for SINR distribution and the convergence speed are shown in Fig.~\ref{SINR_uniform} and~\ref{Iterations_uniform}, respectively. It should be noted that the utility function in Fig.~\ref{Iterations_uniform} is equal to $U_1$ if $U_1$ is greater than zero; for negative values of $U_1$ and also if $U_2$ is greater than $-\zeta N_U$, the utility function is equal to $U_2$; otherwise, it equals $U_3$.\\
\indent In Scenario II, 40 percent of the users are normally distributed with standard deviation 1000 meters in the central region. The other 60 percent of the users are uniformly distributed in the remaining region. In this scenario, the final number of required drone-BSs to serve all the users is 29. The user distribution and the 2D projection of drone-BSs locations is depicted in Fig.~\ref{BSC_deploy_normal}. As seen in this figure, in the central region which has higher user density, the drone-BSs are closer to each other compared to the other subarea. Also like previous scenario, in the region with higher user density, the drone-BSs are placed in lower altitudes to make less interference for the users served by other drone-BSs. Fig.~\ref{SINR_normal} and~\ref{Iterations_normal} show the CDF curve for SINR distribution and the convergence speed of the algorithm in Scenario II, respectively.

\section{Conclusion}
\indent This paper provided a new drone-BSs deployment plan, while minimizing the number of them, in order to serve the users based on their traffic requirements. Generally this is an optimization problem which is too complex to solve, therefore a heuristic algorithm based on particle swarm optimization was proposed. The number of drone-BSs and their 3D placement were estimated, while satisfying coverage and capacity constraints of the system. Afterwards, the drone-BSs whose removal did not affect the quality of the network, were removed. Simulation results considering regions with different user densities, confirmed the acceptable performance of the proposed method. It was seen that the number of drone-BSs in an area is proportional to the user density in that area. It was also noted that drone-BSs can change their altitudes in order to tackle coverage and capacity issues. A drone-BS decreases its altitude in a dense area to reduce interference to the users that are not served by it and increases its altitude to cover a large area in a low density region.

\bibliographystyle{IEEEtran}
\bibliography{IEEEabrv,xbib}

\end{document}